# NASA'S GREAT OBSERVATORIES – A TRIUMPH OF THE HUMAN SPIRIT

[proposed revision, 2 August 2025]

Michael Werner – mwwerner800@gmail.com

Jet Propulsion Laboratory, California Institute of Technology

**ABSTRACT** In January of 1985, more than 40 years ago, a group of astronomers met with NASA officials to map out the future of NASA space astronomy. Their efforts led to the Great Observatories program, linking four powerful space telescopes to study the heavens in four regions of the spectrum. The successful launch and operation of the Spitzer Space Telescope in the Fall of 2003 completed the launch of the Great Observatories, almost 20 years after the program was formulated, and two of the Observatories, Hubble and Chandra, continue to operate very productively. The scientific and public education results of the Great Observatories are well-known. Here we emphasize that fulfilling the extraordinary vision of the Great Observatories was a triumph of human ingenuity, dedication, and determination.

## 1. THE ROOTS OF THE GREAT OBSERVATORIES

People have studied the heavens for millennia, but until the 1930's the human eye, and telescopes operating in the visible band, were the only tools available for these studies. Technical developments during the 20th Century revolutionized astronomers' view of the heavens, starting in the 1930's with the birth of radio astronomy, and continuing with the advent of telescopes operating throughout the electromagnetic spectrum. During the 1960's space-based explorations of the Universe in the ultraviolet, Xray, and Gamma Ray bands allowed astronomers to see light that could not get through the atmosphere (Figure 1). By the 1980's, NASA had plans for launching major facilities covering each of these spectral regions. These plans also included the visible band, as a telescope in space does not suffer the blurring of images by Earth's atmosphere, and the infrared band, where sensitive observations require a cold telescope in space, as discussed below.

## 2. THE RATIONALE FOR THE GREAT OBSERVATORIES

The scientific underpinning of the Great Observatories program is that each telescope, operating in a different spectral band, would uniquely sample different astrophysical processes, and that all spectral bands are needed to probe the full range of astronomical phenomena. In addition to each having its own unique region of the spectrum, the Observatories often observe synergistically to provide complementary views of the same object.

Figure 1 shows that each spectral band corresponds to a particular range of temperatures for the phenomena under study, as required by basic physics. Temperatures are given in K, or degrees Kelvin, which is the temperature above absolute zero; for comparison, the freezing point of water is at 273K.

The four members of the Great Observatories family include:

**The Compton Gamma Ray Observatory, CGRO,** launched in 1991 and brought down from orbit in 2000 to avoid an uncontrolled reentry into the Earth's atmosphere. Gamma Rays sample very energetic phenomena, often occurring in the vicinity of black holes. Work on Compton began prior to 1977, at least 14 years before launch.

**The Chandra Xray Observatory, CXO**, known prior to launch as AXAF. Chandra launched in 1999 and is still operating. As one example, Xrays sample hot plasmas associated with the remnants of exploded stars. Work on AXAF/CXO began in 1977, or 22 years before launch.

Th**e Hubble Space Telescope, HST**, launched in 1991, is still operating. thanks to five successful Space Shuttle based servicing missions. Hubble covers the ultraviolet and visible bands, and even a small slice of the infrared. Its science reach extends from studies of stars to exoplanets – planets around nearby stars – out into the distant Universe. Work leading to Hubble started in 1968, or 23 years before launch.

T**he Spitzer Space Telescope, referred to prior to launch as SIRTF,** operated throughout the infrared band. Science themes for Spitzer are presented below. Work on SIRTF/Spitzer began in 1972, 31 years before launch.

The start dates cited above correspond roughly to when NASA began to support serious studies that led to each of the Great Observatories. Note that the mean time from the start of the studies to launch is more than two decades. The time was spent in technology development, science and instrument studies, defining and redefining the mission as the external environment changed, etc. Efforts are underway to reduce this time for future missions, but the fact that many folks stayed with the programs during this entire time exemplifies the dedication of those who have worked on these Observatories.

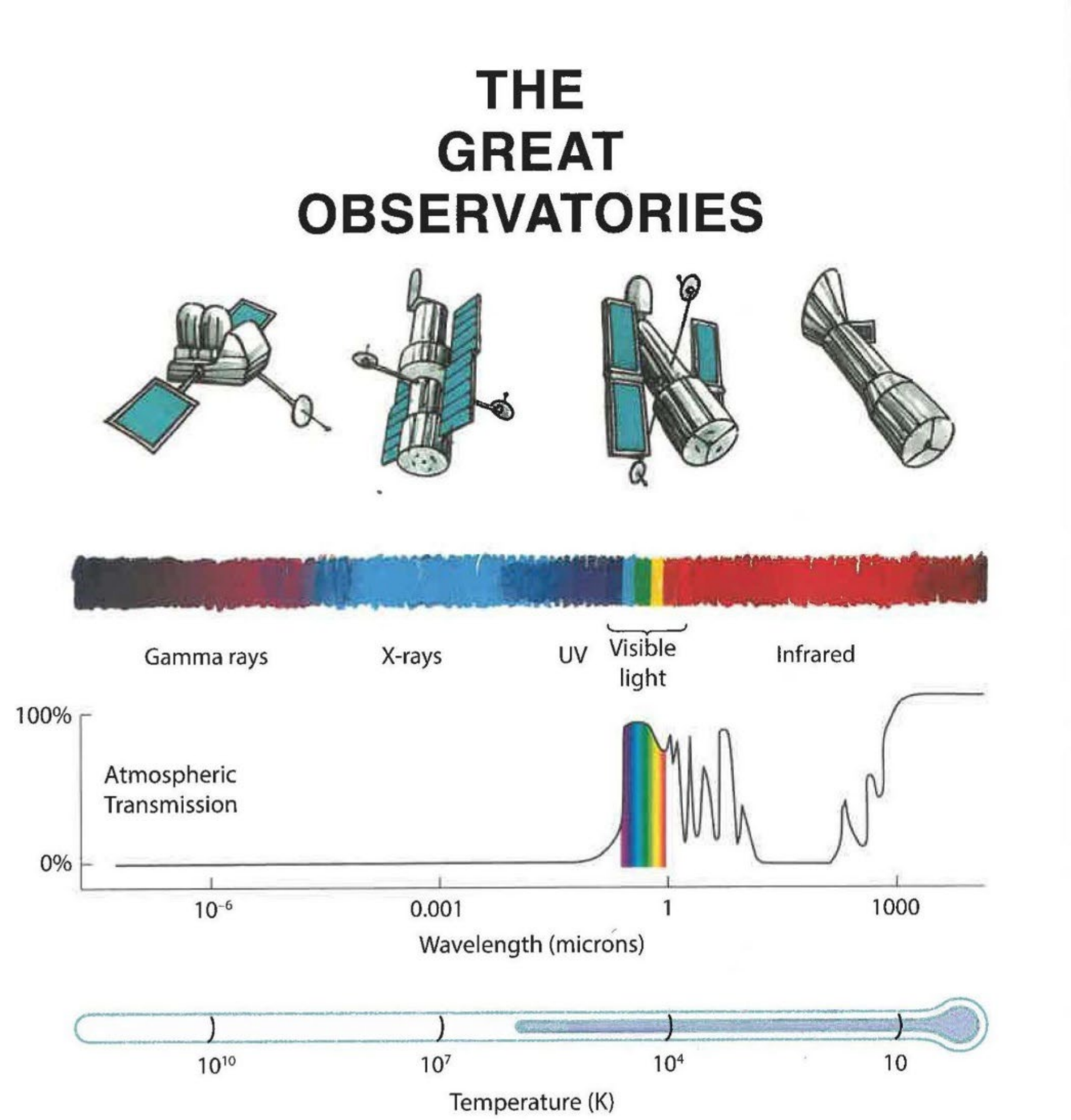

## Figure 1

This image is adapted from the Great Observatories brochure. It shows, from left to right, the four Great Observatories: Compton, Chandra, Hubble, and Spitzer [not to scale], each with its unique spectral band. Also shown, below, are the transmission of the Earth's atmosphere across the spectrum, and also the temperature and wavelength region uniquely associated with each spectral band.

## 3. THE BIRTH OF THE GREAT OBSERVATORIES

Regardless of the scientific imperative, it was as much funding concerns as science that motivated the formulation of the Great Observatories. Martin Harwit of Cornell University was a prime scientific force in this formulation; he has described the origins and early history of the Great Observatories in several publications (1,2):

In late 1984 Charlie Pellerin, then NASA's Division Director for Astrophysics, realized that his cup was overflowing. He had two major missions, Compton and Hubble, under active development; they had been approved by Congress to go forward. Two others, Spitzer and Chandra, were undergoing significant studies, supported by highly vocal advocates within the scientific community (even though Chandra and Spitzer were known, respectively, as AXAF and SIRTF until they were up and running, we refer to them throughout as Chandra and Spitzer) At the same time, NASA was seeking support within the Administration and Congress for the development of the International Space Station.

Advocates for both Spitzer and Chandra were separately visiting Charlie's office to argue that their mission should have higher priority and move forward. In Charlie's own words (1):

"The argument might have been worth having if one [of Spitzer and Chandra] supplanted the need for the other. In fact, either mission increased the necessity for the other.... I needed to create a fresh, compelling strategy that would mobilize everyone behind both missions and the two already under way [Hubble and Compton]."

Charlie responded in classic NASA fashion: he called the meeting described earlier. In terms of impact and significance, this was no ordinary NASA meeting. The attendees, both from academia and from NASA centers, included leaders in many areas of both experimental and theoretical astrophysics.

According to Harwit (1,2), who chaired the meeting, the main purpose was to prepare a booklet which would help Charlie and others to explain the astrophysics program. Starting with some sketches which Harwit had prepared, the attendees drafted graphics to show how the observatories would work together to attack major astrophysical problems. At the end of the meeting, the draft sketches were given to science historian Valerie Neal, who worked with Harwit and artist Brien O'Brien, to produce the final booklet. In colorful comic book style, the final booklet argues persuasively for the importance of a family of observatories in space (Figure 2) and for the technologies required to bring them to life. As the booklet neared completion, George Field of Harvard suggested calling the program "The Great Observatories", which captured its spirit brilliantly (1).

The Great Observatories was an early example of the multimission, multiwavelength perspective which motivates much of current astrophysical research. Radio astronomy was excluded because it can be done from the ground, although the complementarity of radio observations to the scientific programs of the Great Observatories was frequently acknowledged [see Figure 2]. An updated version of the booklet might also include the contributions of neutrino and gravitational wave astronomy, newly burgeoning fields of study which started only after the formulation of the Great Observatories.

Figure 2 [below] conveys the colorful and accessible style of the Great Observatories booklet. It was immediately effective, as tens of thousands of copies were delivered to astronomers, the general public, and government decision makers. Once the idea of the Great Observatories took hold, advocates for Chandra and Spitzer could argue that they completed a program which had been started with the earlier selection of Compton and Hubble. Marcia Rieke of the University of Arizona, who spearheaded advocacy for Spitzer in Congress very capably, reported:

"Being part of the Great Observatories [GO] was very helpful as one of the usual questions was 'why do you need another space telescope?' The GO presented a nice framework for explaining why another one was needed." (M.Rieke, private communication).

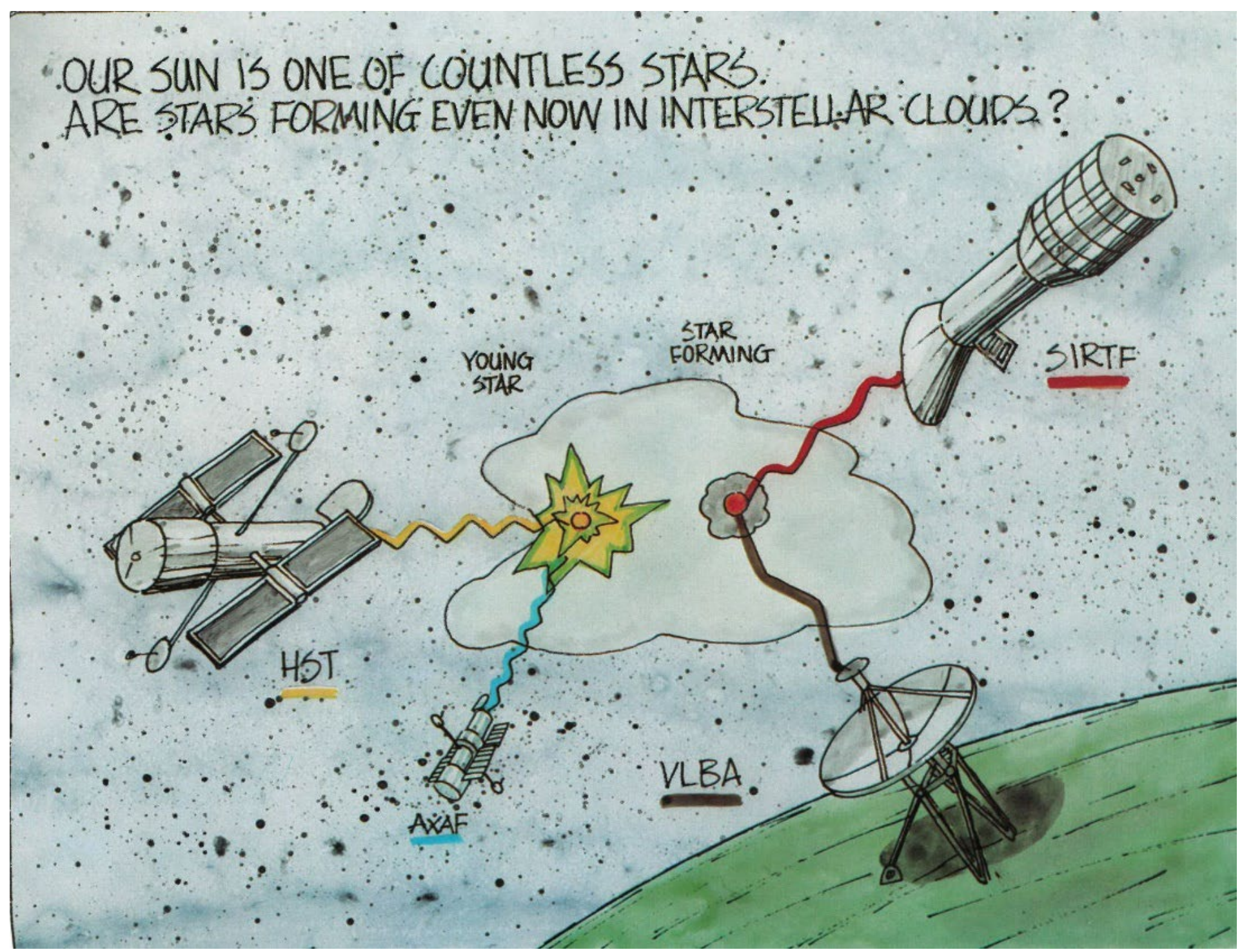

Figure 2. A page from the Great Observatories booklet showing how HST (Hubble), AXAF (Chandra), and SIRTF (Spitzer) would work together to address "How are Stars Born?" one of the major scientific themes presented in the booklet. The VLBA, an array of ground-based radio telescopes, is included in the image for scientific completeness. The booklet has been affectionately referred to as "The Great Observatories Coloring Book [or Comic Book]".

Chandra and then Spitzer were approved for final design and construction after the release of the Great Observatories booklet.

## 4. THE IMPACTS OF THE GREAT OBSERVATORIES

The impacts of the Great Observatories have been profound:

**A. Scientific** The Observatories have been tremendously important scientifically. Scientific publications using data from Hubble, Chandra, and/or Spitzer amounted to about 1800 papers per year, for a total of 5400 papers over the three years 2010-2012, most of which had multiple authors. In 2010, the membership of the American Astronomical Society stood at 6,680, suggesting that at least 50% of the US astronomical community participated in the scientific bounty of the Great Observatories during those years.

Much of the science executed by Hubble, Chandra, and Spitzer was carried out by scientists with no previous involvement with that mission. (Compton was something of a hybrid in this respect and was not as widely used as the other Observatories.) The Hubble, Chandra, and Spitzer teams established dedicated science centers to facilitate the work of the community. International contributions varied among the Observatories, with HST, in particular, being a strong partnership with the European Spae Agency. In all cases, however, proposals from international scientists were encouraged, and selected proposals with supported the funds from the proposers' home agencies. With support from NASA, the science centers invested heavily in educational material and hands-on teacher training activities which have extended their impact throughout the US and across the world.

The Great Observatories is a gift that keeps on giving. Even after Hubble and Chandra join Spitzer and Compton in retirement, the Observatories will live on because all data are archived and readily available to scientists and non-scientists. Even today, with Chandra and Hubble still operating, a large fraction of the publications from these Great Observatories are based on archival data.

The scientific bonanza of the Observatories is too vast to be dealt with in the present article. Detailed discussion of the science of the Great Observatories appear elsewhere [3,4,5,6].

**B. Technical** In many cases the Great Observatories pioneered the use of technologies in space which were then adopted by future missions. Two examples which come to mind immediately are the adoption and demonstration of extensive radiative cooling by Spitzer and the pioneering use of CCDs by Hubble.

**C. Cultural** Although they have rewritten the textbooks, the impact of the Great Observatories goes far beyond professional and formal educational circles. Starting with the initial repair of Hubble's optical system , the press and the public have been blessed with fabulous images and exciting scientific insights from the Great Observatories. These releases have fallen on fertile soil, as many members of the general public are fascinated by space amd astronomy. In addition, images from the Great Observatories find their way to mugs, t-shirts, calendars, and all manner of commercial products. Images from the Great Observatories appear throughout the world. For example, images from Spitzer have appeared on the fence at the Jardin du Luxembourg in Paris.

**D. Human** The completion of the launch and initial operation of the Great Observatories with the successful launch of Spitzer, almost 20 years after the program was first proposed, is a great human achievement, which has been likened to building a medieval cathedral [7].

At an informal meeting in the early 1970's, Nobel-prize winning physicist Charles Townes pointed out that science is not a zero-sum game. The success of the Great Observatories stands as a tribute to the vision, spirit, dedication, determination, and ingenuity of the tens of thousands of scientists, engineers, technicians, managers, and support staff - who have worked on the creation and operation of the Observatories, often for two decades or more. Each member of this multitude should be proud of, and recognized for, their contributions to a historically successful and impactful program.

## 5. THE SPITZER SPACE TELESCOPE AND THE IMPACTS OF INDIVIDUALS

**A. What a Single Person Can Do** The saga of the Great Observatories demonstrates another important and inspirational principle by showing that, even in a large, expensive, complex project, a single indivdual in the right place at the right time can have a major impact. We discuss below the importance of individuals in the development of the Spitzer Space Telescope. Histories of Spitzer which include the issues raised below and much more have been presented by Rieke [8] and Rottner [9]. Spitzer is not unique in this respect; similar examples could be found in the stories of the other Great Observatories [3,5,6].

**B. A Prime Directive** The prime directive in the design and operation of Spitzer was to keep it cold - just a few degrees above absolute zero. Figure 1 shows that objects at around the Earth's temperature, 300K or so, radiate very effectively in the infrared band. This infrared radiation from an earth-based telescope, and from the atmosphere itself, can overwhelm the signals from faint celestial targets Visible wavelengths present a useful analogy. The bright sunlit atmosphere hides the faint stars, which can readily be seen at night once the sun is down. But the temperatures of the telescope and the atmosphere, and thus their infrared radiation, do not decrease appreciably from day to night. **So to work at night – with the highest possible sensitivity - in the infrared requires a cold telescope above the atmosphere**, where the brightness of the sky in the infrared is millions of times fainter than encountered from a warm telescope within the atmosphere, and objects hidden swamped by the bright skies can become readily visible. Then, as was demonstrated by Spitzer's precursor mission IRAS in 1983, and is being demonstrated dramatically by the recently launched James Webb Space Telescope, the heavens come alive with sources of infrared radiation, going far beyond the important results reported by pioneering observers using Earth-based telescopes.

**C. A Tale of Two Spitzers**

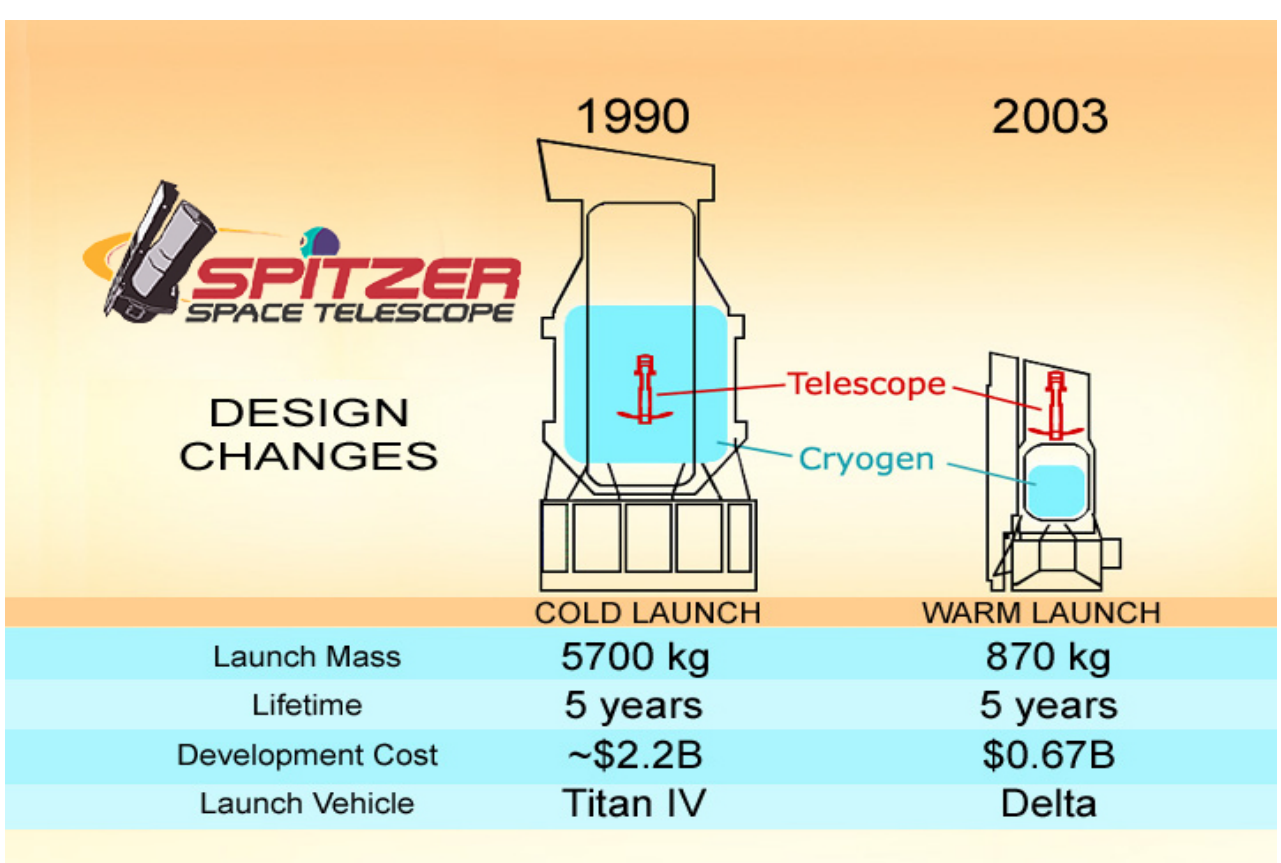

| | COLD LAUNCH | WARM LAUNCH |
|---|---|---|
| Launch Mass | 5700 kg | 870 kg |
| Lifetime | 5 years | 5 years |
| Development Cost | ~$2.2B | $0.67B |
| Launch Vehicle | Titan IV | Delta |

Figure 3: Two versions of the Spitzer Space Telescope, the Cold Launch version of 1990, and the Warm Launch system actually launched in 2003. The blue color represents the liquid helium coolant, at a temperature of just a degree or two above absolute zero, used to cool or control the temperature of the telescope and instruments. The lifetime refers to the lifetime of the liquid helium supply. Figures such as this were used as examples of NASA administrator Daniel Goldin's "Faster, Better, Cheaper" approach to NASA science missions (See ch.7 of Ref.3).

Figure 3 shows, to scale, two versions of the Spitzer Space Telescope. The "Cold Launch" version on the left was selected by the once per decade review of Astronomy and Astrophysics (chaired by John Bahcall of Princeton) as the highest priority major space astronomy project for the 1990's. The "Warm Launch" version shown on the right shows Spitzer as it was launched in 2003. The two versions have comparable telescope size and cryogen lifetime, and thus similar scientific power, but the Warm Launch Spitzer was much smaller and less expensive than its predecessor.

The ink had barely dried on the Bahcall Report when problems with the Hubble Space Telescope primary mirror, the loss of NASA's Mars Observer spacecraft, and the antenna issues with the Galileo mission to Jupiter ruled out billion+$ missions like the Cold Launch, and eventually Spitzer was subjected to a cost cap of $500M. The Spitzer team responded effectively to this new challenge in just a few years by developing the Warm Launch mission shown above.

## 6. GETTING FROM COLD TO WARM

It is the author's opinion that no amount of tinkering or optimization of the Cold Launch system would have led to the Warm Launch Spitzer. In retrospect, the Spitzer team went back to square one and started off towards the Warm Launch. The path from the Cold to the Warm Launch was navigated by three paradigm-breaking contributions, proposed by specific individuals at specific times. These are described briefly below; the first two are discussed in greater detail In the Appendix. Further technical details concerning the on-orbit performance of Spitzer are given by Werner et al (11,12).

**A. A Kinder, Gentler Orbit** At a meeting of the Spitzer Science Working Group (which had been selected by NASA in 1983 to provide scientific advice and guidance to the Spitzer project), in the Spring of 1992 JPL Mission Engineer Johnny Kwok pointed out that giving the spacecraft just enough energy to escape Earth's gravity and to go into orbit around the sun, had numerous avantages for Spitzer. Spitzer would follow the Earth around the sun, drifting further behind the Earth by about ten million miles per year. The advantages of this approach, as described by Kwok et al (13 ) and summarized in the Appendix, were so obvious and immediate that the Science Working

Group adopted it enthusiastically. Subsequent engineering analysis validated their enthusiasm and this "drift-away orbit" became the choice for Spitzer in record time. Johnny's suggestion was paradigm breaking because all previous NASA astrophysics missions – and versions of Spitzer - had been in Earth orbit.

**B Using the Refrigerator of Space.** In the Fall of 1993 the Spitzer SWG met at a Ball Aerospace facility in Broomfield, CO, to chart a path for Spitzer through the murky programmatic waters which confronted us. On the second morning of this retreat, Facility Scientist Frank Low of the University of Arizona came down to breakfast with a sketch proposing the Warm Launch architecture for Spitzer. Frank pointed out that in the Warm Launch approach, the cooling of the telescope and instruments would be achieved and maintained largely by radiation into the cold refrigerator of deep space rather than by evaporation of stored liquid helium cryogen, as was the case for the Cold Launch and for previous space missions. This approach is called "radiative cooling". The radiative cooling brought the telescope and instruments down to an operating temperature ~40K, and a small amount of liquid helium from the cryogen tank provided the final cooinling to temperatures close to absolute zero. Again, detailed engineering analysis verified this approach, with its many technical advantages, and like the drift-away orbit to which it was particularly well suited [Appendix A], it, too, became part of the Spitzer mission concept going forward.

Prior to Frank's suggestion, a group led by Tim Hawarden [had already proposed *POIROT*, a mission which relied on radiative cooling, to a European Space Agency call for proposals. (The early development of radiatively-cooled telescopes is well-documented by Harley Thronson (10), who led the definition of *Edison*, which further advanced this acrchitecture). Nevertheless, Frank's suggestion was paradigm breaking, because for more than two decades work on Spitzer had considered only the Cold Launch approach, which had been used with great success in IRAS and in the European Space Agency's highly successful Infrared Space Observatory (ISO) mission.

**C. An Important Collataeral Benefit** As outlined in the Appendix, the drift-away orbit and the warm launch played together very well. This allowed the initial supply of liquid helium, which had maintained the telescope and the payload at temperatures approaching absolute zero, to last for almost six years. Importantly, after the helium was exhausted, the telescope and the instruments warmed up to only around 30 degrees above absolute zero. At this still very low temperature, several of Spitzer's detectors were able to continue operating without loss of sensitivity. This so-called "Warm Spitzer" phase continued to return important scientific results, particularly in the burgeoning field of exoplanet studies, for an additional 12 years, until January 2020.

**D.Where Spitzer shines** At the Broomfield retreat mentioned in **B** above, Spitzer science team members Mike Jura and George Rieke proposed to limit the advertised Spitzer science case to four important scientific thrusts, traceable to the Bahcall report which had given Spitzer the nod in 1990. The team adopted this idea enthusiastically, reasoning correctly that a system which was designed to do a few things superbly could do many other types of science very well. Jura and Rieke's suggestion was paradigm-breaking because Spitzer had previously publicized a smorgasboard of scientific objectives, intending to be all things to all astronomers. When limiting the advertised scope of Spitzer science was adopted, it included the idea that only instrumental capabilities necessary to study the adopted themes would be included in the payload. This led to a reduction in the size and mass of the instrument package which brought it into balance with the other characteristics of the Warm Launch.

The four themes adopted for Spitzer were:

1. **Protoplanetary and planetary debris disks.**
2. **Brown dwarfs and superplanets.**
3. **Ultraluminous galaxies and active galactic nuclei.**
4. **The distant Universe**.

Spitzer made great strides addressing these themes, and many others which were never envisioned by the designers of the system (4). The scientific bounty of Spitzer was possible because of the superb technical performance of the Observatory, the ingenuity of the scientific community, the use of radiative cooling in the solar orbit, and its unexpectedly long lifetime.

It is interesting to trace the impact of these individual contributions on missions which followed Spitzer. The driftaway orbit was adopted for NASA's revolutionary Kepler exoplanet Discovery Mission, but many of the more recent missions, such as the James Webb Space Telescope (JWST), have been launched into the L2 Lagrange point of the Earth-Sun system; L2 is located approximately one million miles on the opposite side of the Earth from the Sun. L2 has all of the advantages of the driftaway orbit, and, in addition, permits a high and unchanging telemetry rate for sending data back to Earth, unlike the driftaway orbit where the Spitzer-Earth distance increased with time to the point where communications with the satellite became prohibitively inefficient.

Spitzer's successful demonstration of radiative cooling led in part to the radiative cooling adopted by NASA missions large (JWST) and small (SPHEREx). Finally, limiting the advertised science goals of a mission, as was done for Spitzer, has now become the new normal. JWST (4 themes) , the recently-launched SPHEREx Explorer (3 themes), and the far infrared probe PRIMA (3 themes), currently under study, have based their science case and their performance requirements on a small number of themes.

**7.CONCLUSIONS** The Great Observatories live on, although the program was proposed more than 40 years ago. Their science, whether contemporary or archival, continues to be of the greatest significance, and the Observatories have surely exceeded any expectations. The completion and use of the Great Observatories is a monument to the contributions of tens of thousands of folks over the years, while at the same time, as we have shown in the discussion of Spitzer, there have been times when the contribution of a single person has been critical in moving the program forward. The completion and success of The Great Observatories provide a sterling example of the very best that a group of talented people, highly motivated and empowered, can achieve.

## 8.REFERENCES

## 9. ACKNOWLEDGEMENTS

The author had the great good fortune to serve as Spitzer Project Scientist for more than 30 years, so that he witnessed the birth of the Great Observatories and participated in the Spitzer-related activities described in the latter portion of this paper. He thanks Martin Harwit, Paul Hertz, George and Marcia Rieke, Tom Soifer, and Martin Weisskopf for useful comments, Harley Thronson for clarifying this early history of concepts for radiatively-cooled telescopes, and Claire Gorfinkel for editorial assistance.  He gratefully acknowledges the foresight of Martin Harwit and Charlie Pellerin in bringing the Great Observatories to life more than 40 years ago.   That history provides another dramatic example of what one person [or two in this case] can do.

The research was carried out at the Jet Propulsion Laboratory, California Institute of Technology, under a contract with the National Aeronautics and Space Administration(80NM0018D0004).

**APPENDIX.  The Technical Motivation for the Warm Launch Spitzer.**

If Spitzer orbited the Earth, as was to be the case for the Cold Launch system, it would not have been possible to carry out a sequence of observations without the light from the Earh striking and heating the outer shell of the telescope.  (The outer shell is the roughly cylindrical structure which surrounds the cryogen tank, the instruments, and the telescope in both versions of Spitzer shown in Figure 3). This heat input, together with the heat that leaked inward from the solar panel,   (Although not apparent in Figure 3, in each concept a solar panel converts incident sunlight into electrical power and shields the outer shell from direct sunlight), would prevent the outer shell from becoming as cold as in the Solar Orbit, when the Earth is much further away, and the heat input from the Earth is no longer an issue.  The Solar Orbit also allowed for easier scheduling of observations and longer observation times, as it was not necessary to avoid the Earth while pointing around the sky.

The Solar Orbit played very nicely with the subsequently proposed warm launch architecture for Spitzer, as it allows the spacecraft to be continually oriented with the solar panel close to perpendicular to the incoming solar radiation, making it possible to optimize the novel, radiatively cooled architecture with no concerns about sunlight, or light from the now very distant Earth, striking directly any other surface of the observatory.  In fact, the half of the outer shell not facing the solar array  was painted black to enhance the radiative cooling intrinsic to the Wam Launch.

In this Warm Launch configuration, the telescope was launched warm at around 300K and cooled down on orbit by radiating into the cold void of space.

This eliminated the large cryostat required for the Cold Launch Spitzer.  The outer shell is smaller and colder [35 K vs. 110 K], due in part to the absence of light from the Earth striking the outer shell, and also to the power of the radiative cooling, which operates continuously

In the Warm Launch case, the Observatory carried only enough liquid helium to keep the instruments cold and to bring the telescope down to its operating temperature of ~5K after the radiative cooling became ineffective around 40K. This required only 350 liters of liquid helium at launch rather than the 4000 liters required for the Cold Launch, where the heat percolating inward from the warm and large outer shell would have reached the helium tank and vaporized the liquid helium, just as a kettle on the stove boils water.   This heat input was negligible for the Warm Launch system because of the smaller and much colder outer shell  The liquid helium was heated only by the small amount of power needed by the instruments and boiled away much more slowly.

The larger helium tank would have had implications for the rest of the Cold Launch Spitzer.  The cryostat [which contains the liquird helium tank and the instrument package in each architecture] and the outer shell would be larger for the Cold Launch than was flown in the Warm Launch case  to accommodate the larger helium tank and to fit the telescope into the cryostat. Further increases in relative cost would have come from other elements of the system; a much larger spacecraft would be needed to orient the system, and, significantly, a much larger and costlier rocket would have been needed to launch the Cold Launch Spitzer, although this impact is not included in the cost differential shown in Figure 3.